\documentclass[journal=nalefd,manuscript=letter]{achemso}
\pdfoutput=1
\usepackage[version=3]{mhchem} 

\usepackage{dcolumn}
\usepackage{bm}

\usepackage{epstopdf}
\usepackage{graphicx}
\usepackage{amsmath,amssymb}

\author{Samuel Berweger}
\email{samuel.berweger@nist.gov}
\affiliation[NIST Boulder]
{Applied Physics Division, National Institute of Standards and Technology, Boulder, CO}
\alsoaffiliation[CU Physics]
{Department of Physics, University of Colorado, Boulder, CO}
\author{Gang Qiu}
\affiliation[Purdue ECE]
{School of Electrical and Computer Engineering, Purdue University, West Lafayette, IN}
\alsoaffiliation[Birck NT]
{Birck Nanotechnology Center, Purdue University, West Lafayette, IN}
\author{Yixiu Wang}
\affiliation[Purdue IE]
{School of Industrial Engineering, Purdue University, West Lafayette, IN}
\author{Benjamin Pollard}
\affiliation[CU Physics]
{Department of Physics, University of Colorado, Boulder, CO}
\author{Kristen L. Genter}
\affiliation[NIST Boulder]
{Applied Physics Division, National Institute of Standards and Technology, Boulder, CO}
\alsoaffiliation[CU ME]
{Department of Mechanical Engineering, University of Colorado, Boulder, CO}
\author{Robert Tyrell-Ead}
\author{T. Mitch Wallis}
\affiliation[NIST Boulder]
{Applied Physics Division, National Institute of Standards and Technology, Boulder, CO}
\author{Wenzhuo Wu}
\affiliation[Purdue IE]
{School of Industrial Engineering, Purdue University, West Lafayette, IN}
\alsoaffiliation[Birck NT]
{Birck Nanotechnology Center, Purdue University, West Lafayette, IN}
\author{Peide D. Ye}
\affiliation[Purdue ECE]
{School of Electrical and Computer Engineering, Purdue University, West Lafayette, IN}
\alsoaffiliation[Birck NT]
{Birck Nanotechnology Center, Purdue University, West Lafayette, IN}
\author{Pavel Kabos}
\affiliation[NIST Boulder]
{Applied Physics Division, National Institute of Standards and Technology, Boulder, CO}

\title{Imaging Carrier Inhomogeneities in Ambipolar Tellurene Field Effect Transistors}

\begin{document}
 
\begin{abstract}

Developing van der Waals homojunction devices requires materials with narrow bandgaps and simultaneously high hole and electron mobilities for bipolar transport, as well as methods to image and study spatial variations in carrier type and associated conductivity with nanometer spatial resolution.
Here we demonstrate the general capability of near-field scanning microwave microscopy (SMM) to image and study the local carrier type and associated conductivity \emph{in operando} by studying ambiploar field effect transistors (FETs) of the 1D vdW material tellurium in 2D form.
To quantitatively understand electronic variations across the device, we produce nanometer resolved maps of the local carrier equivalence backgate voltage.
We show that the global device conductivity minimum determined from transport measurements does not arise from uniform carrier neutrality, but rather from the continued coexistence of \emph{p}-type regions at the device edge and \emph{n}-type regions in the interior of our micron-scale devices.
This work both underscores and addresses the need to image and understand spatial variations in the electronic properties of nanoscale devices.

\end{abstract}

\section*{Introduction}

Developing fundamental circuit building blocks such as diodes and transistors \cite{sze} based on low-dimensional van der Waals (vdW) materials requires producing and controlling adjoining regions of \emph{p}-type and \emph{n}-type transport \cite{ross14,pospischil14,baugher14,buscema14,duan14,liu16}.
While the broad library or vdW materials offers a range of novel functionalities such as memristors \cite{sangwan18}, topologically protected states \cite{kou14,ju15,zhang09}, or those based on magnetic \cite{huang18} and other spin-based phenomena \cite{mak14}, MoS$_2$ and the associated family of transition metal dichalcogenides (TMDs) remain the primary candidates for semiconducting applications.
In these materials bipolar transport for homojunction devices can be difficult to achieve \cite{ross14,chen18}, and difficult-to-fabricate heterostructures remain the primary practical path for pn junctions and other basic semiconducting circuit elements \cite{liu16}.
A key challenge therefore remains the development and discovery of new materials for homojunction devices with the narrow bandgaps and simultaneously high hole and electron mobilities \cite{das14,liu14,wang18} desired for bipolar transport.

Nanometer thin films of elemental tellurium have recently been demonstrated for device applications \cite{wang18}.
Tellurium is a vdW material whose helical 1D atomic chains can be assembled into 1D filaments \cite{mayers02} or 2D films (tellurene) \cite{du17,wang18} and deposited using solution-based processes. 
Tellurene in particular exhibits both \emph{n}- and \emph{p}-type mobilities as high as 700 cm$^2$/V$\cdot$s, and the thickness-dependent bandgap as small as 0.4~eV \cite{wang18} readily enables bipolar transport \cite{qiu18}.
Unlike black phosphorous, another promising material exhibiting bipolar transport \cite{das14,liu14}, tellurene is stable under ambient conditions and resistant to long-term deterioration.
Dopant incorporation is often difficult in 2D vdW materials, and the solution-based synthesis of tellurene further offers an attractive avenue for doping, but successful exploitation of this feature requires a detailed understanding of the resulting local transport properties and their relation to growth conditions.
It is thus of critical importance to be able to address and study not only the local carrier density and associated conductivity in functional devices with nanometer spatial resolution, but to be able to do so in a manner that is also sensitive to the carrier type.

\section*{Results}

\begin{figure}[tb]%%%%%%%%%%%%%%%%%%%%%%%%
\includegraphics[width=\textwidth]{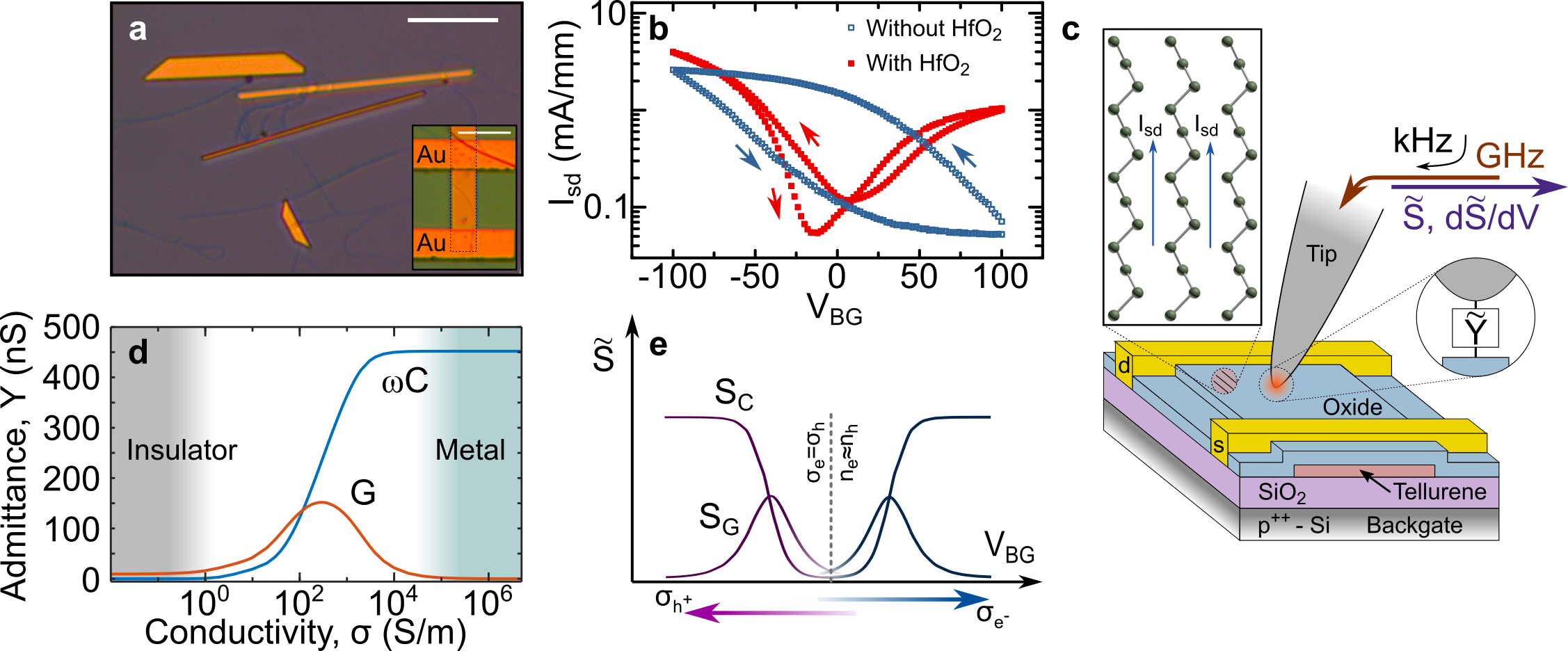}
\caption{
{\bf a} Optical micrograph of solution-grown tellurene crystals showing characteristic trapezoidal shape as well as string-like Te filaments formed during process (scale bar: 20~$\mu$m).
Inset: Optical micrograph of a tellurene device (scale bar: 8 $\mu$m).
{\bf b} Transport characteristics of a tellurene FET before and after deposition of a 10~nm conformal HfO$_2$ overcoat. 
{\bf c} Schematic of the experimental setup illustrating the transport along the direction of the atomic Te chains.
{\bf d} Finite element modeling of the complex-valued tip-sample admittance $\tilde{\rm Y}$~=~G+$i\omega$C at 17.3~GHz as a function of tellurene conductivity.
{\bf e} Illustration of the expected measured microwave signal $\tilde{\rm S} \propto \tilde{\rm Y}$ as a function of V$_{\rm BG}$ for an ambipolar FET showing clear regimes of contrast corresponding to hole- and electron-conductivity. 
}
\label{setup}
\end{figure}%%%%%%%%%%%%%%%%%%%%%%%%

An optical micrograph of representative trapezoidal tellurene flakes deposited on SiO$_2$ is shown in Fig.~\ref{setup}a.
Also seen in the optical micrograph are 1D Te filaments that result as a by-product from solution growth \cite{mayers02}.
Devices are encapsulated in 10 nm of conformal oxide using atomic layer deposition (ALD) of either HfO$_2$ or Al$_2$O$_3$.
Shown in Fig.~\ref{setup}b is the backgate voltage (V$_{\rm BG}$)-dependent transport characteristic of a tellurene FET before and after ALD deposition, acquired with a source-drain bias of V$_{\rm sd}$~=~0.05~V.
In addition to preventing device failure due to tip-sample conduction, the ALD encapsulation also reduces device hysteresis and enables bipolar transport \cite{qiu18}, likely due to oxygen vacancy-induced \emph{n}-doping \cite{valsaraj15}.

An illustration of the experimental setup is shown in Fig.~\ref{setup}c with the scanning microwave microscope (SMM, also called scanning microwave impedance microscopy, sMIM) based on an atomic force microscope (AFM) operating in contact mode that detects the phase-resolved real and imaginary components of the microwave signal $\tilde{\rm S}$~=~S$_{\rm G}$~+~$i$S$_{\rm C}$.
The simulated (COMSOL 4.2*) tip-sample admittance $\tilde{\rm Y}$~=~G~+~$i\omega$C for a unipolar tellurene device is shown in Fig.~\ref{setup}d with the conductance G and capacitance C as a function of the tellurene conductivity $\sigma_i=n_ie\mu_i$ with the elementary charge $e$, carrier density $n_i$, and corresponding carrier mobility $\mu_i$ ($i$~=~$h$, $e$).
Assuming $\tilde{\rm S}$~$\propto$~$\tilde{\rm Y}$ (i.e., S$_{\rm G}$~$\propto$~G and S$_{\rm C}$~$\propto$~C) \cite{berweger15, lai09,wu16}, the SMM sensitivity peaks at $\sigma$~=~300~S/m yet remains sensitive over the range $\sigma$~$\approx$~10$^0$~--~10$^4$~S/m.
Controlling the carrier type and concentration using the global backgate V$_{\rm BG}$, we expect the measured signal $\tilde{\rm S}$ to evolve as schematically illustrated in Fig.~\ref{setup}e with two distinct regions of sensitivity as the device transitions between the hole and electron transport regimes.

\begin{figure}[tb]%%%%%%%%%%%%%%%%%%%%%%%%
\includegraphics[width=\textwidth]{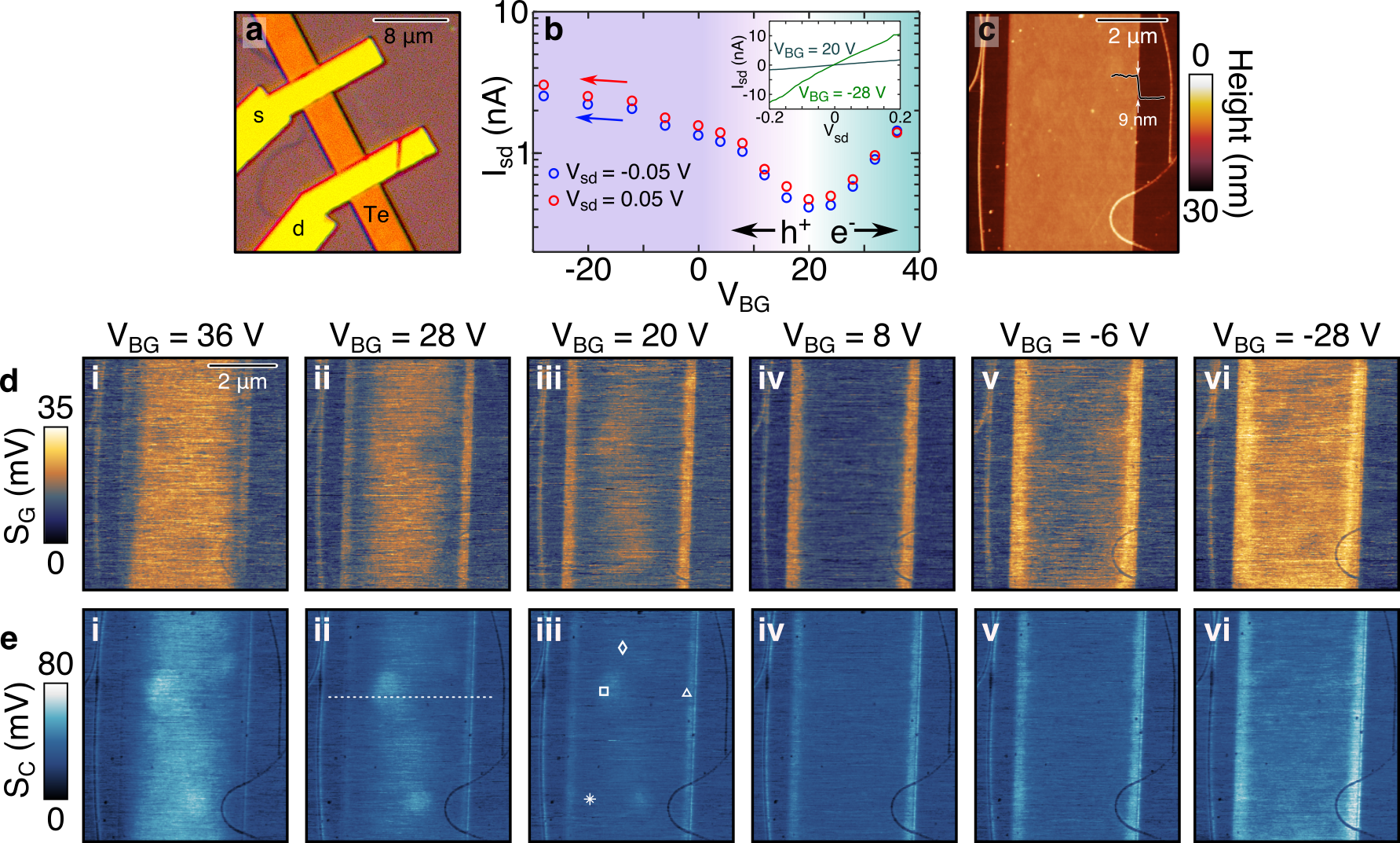}
\caption{
{\bf a} Optical micrograph of an Al$_2$O$_3$-encapsulated tellurene device with source and drain electrodes as indicated.
{\bf b} Transport characteristics of the device, with the global device carrier equivalence point at V$_{\rm BG}$~=~20~V and electron and hole transport regimes as indicated.
Inset: The IV characteristics of the device at V$_{\rm BG}$ as indicated with near-ohmic behavior, though some nonlinearity and increased noise are seen at positive V$_{\rm sd}$.
{\bf c} Contact mode AFM topography of the device active area (source and drain electrodes are just out of view at image top and bottom).
{\bf d} and {\bf e} show SMM images acquired at sequentially decreasing values of V$_{\rm BG}$ as indicated, with the conductance S$_{\rm G}$ and capacitive S$_{\rm C}$ channels, respectively.
}
\label{sequence}
\end{figure}%%%%%%%%%%%%%%%%%%%%%%%%

Shown in Fig.~\ref{sequence}a is an optical micrograph of a tellurene FET, with source and drain electrodes as indicated and a channel length of 8~$\mu$m.
Shown in Fig.~\ref{sequence}b is the source-drain current I$_{\rm sd}$ measured at V$_{\rm sd}$~=~$\pm$0.05~V, showing a clear global conductivity minimum (carrier equivalence) point at V$_{\rm BG}$~=~20~V separating the hole and electron transport regimes as indicated.

The contact-mode AFM topography in Fig.~\ref{sequence}c shows a uniform device with a thickness of $\sim$9~nm and only small topographic variations. 
Also visible are co-deposited Te filaments, including one over the device.
A sequence of S$_{\rm G}$ and S$_{\rm C}$ images at select voltages are shown in Fig.~\ref{sequence}d and e, respectively (see supplement for full data set). 
For large and positive V$_{\rm BG}$ the device is in the electron-transport regime, and the higher signals in both channels indicate that the device interior is more conductive than the device edges. 
As V$_{\rm BG}$ decreases and approaches the global device carrier equivalence point at V$_{\rm BG}$~=~20~V, the interior of the device becomes less conductive while the edge conductivity increases.
When V$_{\rm BG}$ is further decreased and the device transitions into the hole-conduction regime, we see the conductivity in the device interior reach a minimum at V$_{\rm BG}$~=~8~V while the edge conductivity continues to increase monotonically.
As V$_{\rm BG}$ is decreased beyond the conductivity minimum of the device interior, the conductivity in this region is seen to increase again, though over the voltage range accessible with this device it does not reach parity with the edges again. 

Several notable features stand out from the sequence of microwave near-field images.
In particular, the device interior does show spatial conductivity variations.
The filament also affects the measured signal, although this appears to be limited to reducing the overall signal due to the increased tip-sample spacing, and no electronic effects arise in its vicinity.
Most importantly, the discrepancy in V$_{\rm BG}$-dependent behavior between the device interior and exterior suggests inhomogeneities in the carrier type and we perform additional measurements to better understand this behavior.

\begin{figure}[tb]%%%%%%%%%%%%%%%%%%%%%%%%
\includegraphics[width=\textwidth]{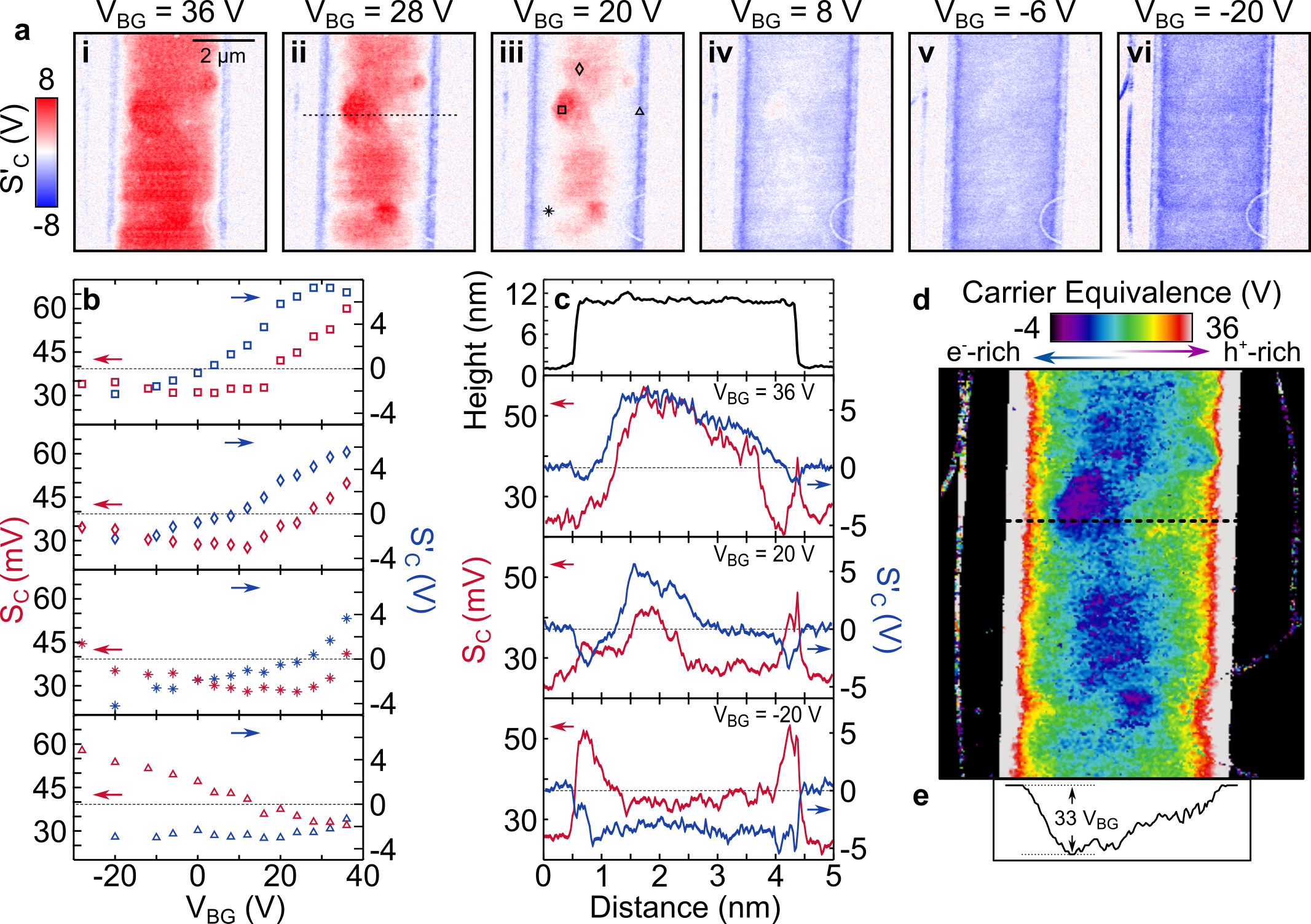}
\caption{
{\bf a} S$'_{\rm C}$ images acquired at V$_{\rm BG}$ as indicated.
{\bf b} S$'_{\rm C}$ and S$_{\rm C}$ as a function of V$_{\rm BG}$ acquired at the locations of the corresponding symbols in aiii and Fig.~2eiii.
{\bf c} Line cuts across the tellurene device showing the topography  (top) and S$'_{\rm C}$ and S$_{\rm C}$ at values of V$_{\rm BG}$ as indicated, taken from the location of the dashed line in aii and Fig.~2eii. 
Dashed lines in b and c are a guide to the eye at S$'_{\rm C}$~=~0~V.
{\bf d} Spatially resolved map of the local carrier equivalence voltage.
{\bf e} Line cut taken along the dashed line in d showing a variation of V$_{\rm BG}$~$>$~33~V across the device.
}
\label{cut}
\end{figure}%%%%%%%%%%%%%%%%%%%%%%%%

Differential measurements are sensitive to the slope of the conductivity-dependent $\tilde{\rm S}$ signal.
Shown in Fig.~\ref{cut}a are the S$'_{\rm C}$~$\equiv$~dC/dV$_{\rm tip}$ \cite{berweger15,huber12} images corresponding to the S$_{\rm C}$ images shown in Fig.~\ref{sequence}d and e (see supplement for full data set, including S$'_{\rm G}$).
The sign of S$'_{\rm C}$ directly reflects the carrier type locally present beneath the tip, and gives a direct measure of the local \emph{p}- or \emph{n}-type character of the tellurene device.
For instance, over a \emph{p}-type region the positive part of the V$_{\rm tip}$ modulation cycle locally depletes (repels) \emph{p}-type carriers and thus reduces the conductivity, giving an out-of-phase response (S$'_{\rm C}$~$<$~0, blue regions). 
For the \emph{n}-type regions the opposite will occur, where the positive portion of the cycle accumulates (attracts) \emph{n}-type carriers and thus increases conductivity, yielding an in-phase response (S$'_{\rm C}$~$>$~0, red regions).

Together with S$_{\rm C}$, S$'_{\rm C}$ provides a more comprehensive understanding of these devices.
Shown in Fig.~\ref{cut}b are the V$_{\rm BG}$-dependent S$_{\rm C}$ and S$'_{\rm C}$ signals at the position of the corresponding symbols in Fig.~\ref{sequence}eiii and Fig.~\ref{cut}aiii.
For the three positions within the device interior, S$_{\rm C}$ decreases with decreasing V$_{\rm BG}$, reaches a minimum value at the local carrier equivalence point, and then increases again, while S$'_{\rm C}$ crosses zero (dashed lines) at the same V$_{\rm BG}$ values  as the minima of S$_{\rm C}$, confirming their differential relationship.
At the device edge, S$_{\rm C}$ increases monotonically with decreasing V$_{\rm BG}$ while S$'_{\rm C}$ remains negative over the full voltage range as expected.

We next take a closer look at the spatial variations in S$'_{\rm C}$ and S$_{\rm C}$ across the tellurene device.
Shown in Fig.~\ref{cut}c are line cuts taken at the position of the dashed line in Fig.~\ref{sequence}eii and Fig.~\ref{cut}cii with the topography (top panel) and corresponding S$'_{\rm C}$ and S$_{\rm C}$ cuts at values of V$_{\rm BG}$ as indicated.
At large positive bias the conductivity is highest in the interior of the crystal where the positive S$'_{\rm C}$ reflects the \emph{n}-type behavior measured by transport while the edges appear weakly conductive with the negative S$'_{\rm C}$ signal revealing \emph{p}-type conductivity well above the global carrier equivalence point.
At the global device carrier equivalence point we see a clear coexistence between \emph{p}-type behavior at the edges and spatially inhomogeneous \emph{n}-type behavior within the device interior.
Comparison with S$_{\rm C}$ reveals that the \emph{p}-type edges are significantly more conductive than the device interior at this point, though over a significantly smaller area.
Once the device has transitioned to uniform \emph{p}-type behavior at V$_{\rm BG}$~=~-20~V, the conductivity profile is inverted with higher conductivity near the edges, though clear inhomogeneities remain.

In order to fully understand the variations in carrier type we construct a nanometer-resolved carrier equivalence voltage map.
To take advantage of the improved contrast of lock-in detection and to avoid uncertainty in the location of the S$_{\rm C}$ minima, we interpolate the V$_{\rm BG}$-dependent zero crossing of S$'_{\rm C}$  to determine the local carrier equivalence point.
Shown in Fig.~\ref{cut}d is a spatially resolved map of the local carrier equivalence voltage at each spatial pixel in S$'_{\rm C}$ as determined from the full sequence of images (see supplement) \cite{pollard14}. 
This carrier equivalence map reveals the complex variations in carrier type and density across the device.
Note that white regions at the device edges have a carrier equivalence point greater than 36~V$_{\rm BG}$.
The line cut along the dashed line in Fig.~\ref{cut}d, shown in Fig.~\ref{cut}e, underscores the degree of spatial inhomogeneity across the device with lateral variation in the local equivalence point of V$_{\rm BG}$~$>$~33~V.

\begin{figure}[tb]%%%%%%%%%%%%%%%%%%%%%%%%
\includegraphics[width=\textwidth]{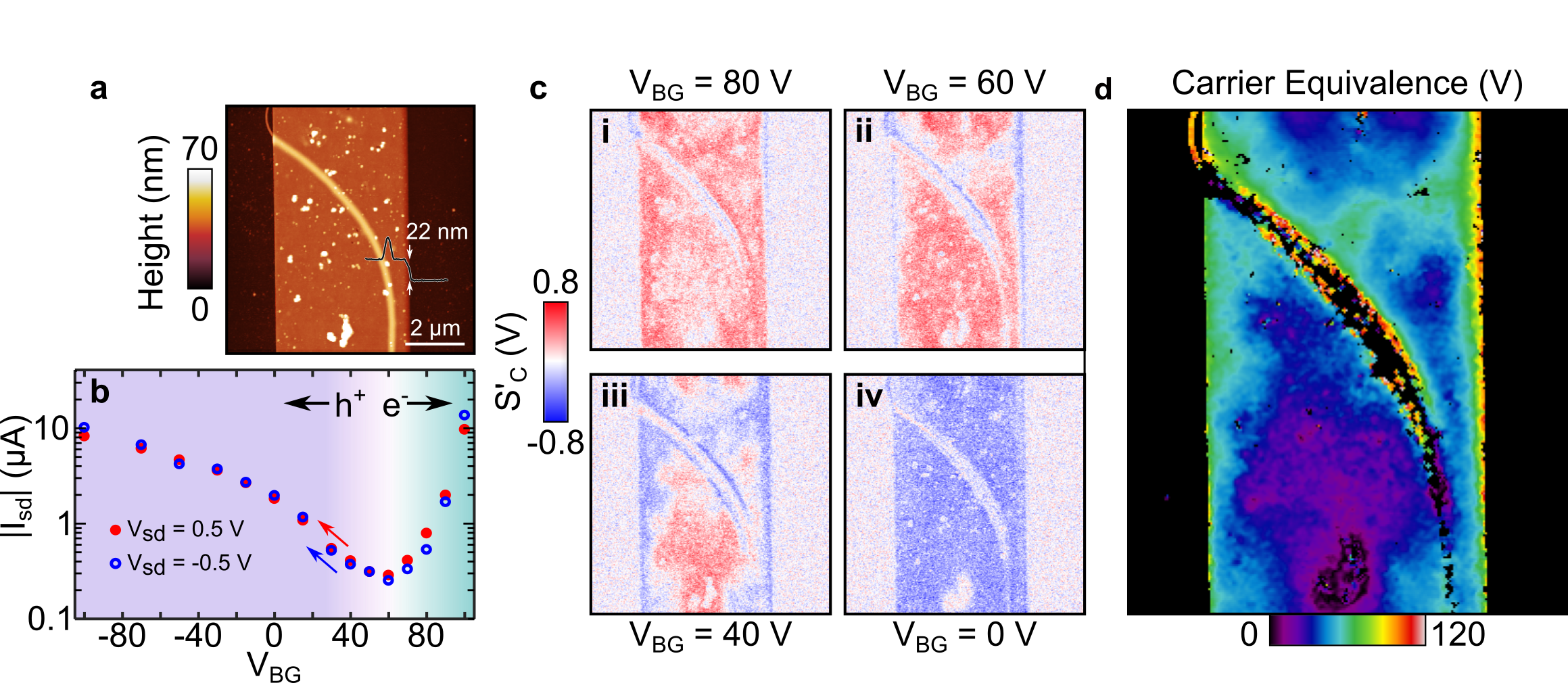}
\caption{
{\bf a} AFM topography of a tellurene device with embedded Te filament.
{\bf b} Transport characteristics of the device with a global carrier equivalence point of V$_{\rm BG}$~=~60~V.
{\bf c} Sequence of S$'_{\rm C}$ images that illustrates the high degree of spatial inhomogeneity in carrier type and density present in this device as further seen in {\bf d}, the extracted carrier equivalence map.
}
\label{device2}
\end{figure}%%%%%%%%%%%%%%%%%%%%%%%%

While thin devices are largely uniform along the transport axis (see supplement for an additional device), large variations in behavior are seen for other devices.
Shown in Fig.~\ref{device2} is the AFM topography of a device where a tellurene filament is embedded beneath the crystal.
This device, encapsulated with HfO$_2$, is deposited on 300~nm of SiO$_2$, which results in a lower backgate efficiency that requires larger voltages up to V$_{\rm BG}$~=~$\pm$100~V, and has a global carrier equivalence point of V$_{\rm BG}$~=~60~V as shown in Fig.~\ref{device2}b.
The S$'_{\rm C}$ sequence in Fig.~\ref{device2}c shows the evolution of the conductivity from \emph{n}-type to \emph{p}-type, with higher hole density at the device edges and an overall electron-rich interior.
However, the embedded filament induces a large degree of spatial inhomogeneity within the crystal, as underscored by the carrier equivalence map shown in Fig.~\ref{device2}d.

\section*{Discussion}

The origin of the spatial variations in conductivity seen across these devices, most notably the significant differences between the edges and interior, is likely associated with several factors.
One of the defining features of vdW materials is the capability of the weak interlayer interactions to accommodate the intercalation of dopants \cite{bediako18,xiong15}, and growth kinetics may favor incorporation of different types or concentrations of \emph{p}-dopants later in the growth process when the edge termination is formed.
We also observe that all images, including the carrier equivalence maps, show local spatial variations in electronic behavior on 10s of nanometer length scales.
These are often correlated with small variations in sample topography (see supplement), suggesting that these variations in electronic behavior are related to the local physical structure resulting from vdW assembly, including molecular alignment, and possibly interfacial effects. 
We note the prevalent strong \emph{p}-type character and associated high local carrier equivalence voltage within the device interior near the embedded Te filament.
This region immediately adjacent to the filament likely experiences significant tensile strain that appears to produce similar electrical characteristics to the edges. 
This suggests that strain strongly influences the carrier density and type.
Lastly, although \emph{n}-type doping due to the ALD overcoat \cite{valsaraj15,qiu18} would be expected to be stronger at the edges where the surface-to-volume ratio is higher, we cannot rule out interfacial impurities and associated fermi-level pinning as contributing factors.

The capability to study and image local variations in conductivity with nanometer spatial resolution has been a long-standing challenge for scanning probe methods.
Optical near-field methods can be used to study conductivity variations in highly doped materials \cite{ritchie17}, and while related near-field photocurrent microscopy has shown potential for graphene \cite{woessner16} its general application has not been explored.
In particular, the differential implementations of SMM and closely related scanning capacitance microscopy (SCM)  have previously been used to study the local carrier type in model systems \cite{edwards00,huber12,berweger15}.
On the other hand, conventional SMM can be used to study nanoscale devices \cite{berweger16,wu16} but has not yet been applied to ambipolar devices.

Studying devices \emph{in operando} requires non-destructive probing in a compatible geometry, and AFM-based methods such as kelvin probe force microscopy \cite{wagner15,koren11} or conductive-AFM \cite{macdonald16} are well suited for such measurements.
However, device encapsulation to prevent destructive electrical discharge from the tip precludes the use of these methods.
This work is thus enabled by leveraging the subsurface imaging capability of SMM \cite{gramse17,wu16} and combining for the first time the conductivity sensing capability of the $\tilde{\rm S}$ signal and the carrier-specific imaging of $\tilde{\rm S}'$ to study bipolar transport in active devices.

Here we have shown how the unique capabilities of microwave near-field microscopy can be used to study the spatial variations in electronic properties of ambiploar field effect transistors. 
In addition to the established capability to image and study spatial variations in sample conductivity, we use differential imaging to determine the local carrier type throughout the backgate voltage-controlled evolution of the device transport properties.
We find large variations in our tellurene devices, with strong \emph{p}-type conductivity observed at the edge of all devices and lower \emph{p}-type conductivity in the device interior that can readily be switched to \emph{n}-type behavior at positive backgate biases.
Our measurements reveal that global transport measurements belie the nanoscale complexity of the device, where we show that the device-average carrier equivalence point is actually the result of the simultaneous coexistence of \emph{p}-type and \emph{n}-type regions that give rise to an overall minimum conductivity rather than true carrier neutrality.
We identify significant variations in the local carrier equivalence point across active devices due to structural variations, interfacial effects, as well as the influence of local strain.
These results underscore the importance of understanding local variations in carrier density and type, and provide a direct means to evaluate the intercalation of dopants into vdW materials.
Our approach can be broadly used to understand and guide the development of low-dimensional semiconducting device architectures and functionalities based on bipolar transport that can be controlled or switched by applied strain or other stressors.

\section*{Methods}

Tellurene devices studied here are fabricated as described previously \cite{wang18}. 
Tellurene films are grown by a solution-based process and deposited using a Langmuir-Blodgett process onto p$^{++}$-Si substrates covered by thermally deposited silicon dioxide, and contact electrodes are subsequently fabricated via electron beam lithography. 
Devices are subsequently encapsulated in 10~nm of conformal oxide (HfO$_2$ or Al$_2$O$_3$) using atomic layer deposition.
The devices are then mounted to the AFM scanner and the leads wirebonded to enable electrical measurements and backgate voltage control \emph{in situ}.

The SMM is based on a modified commercial AFM/SMM system (Keysight*) operating in contact mode \cite{berweger15}.
The microwave signal at $\sim$17.3 GHz is delivered via the tip to the sample using a coaxial resonator and the reflected $\tilde{\rm S}$ signal is detected using an IQ mixer (Analog Devices*).
A phase controlled reference signal is used to separate the real and imaginary components $\tilde{\rm S}$~=~S$_{\rm G}$~+~$i$S$_{\rm C}$.
Differential measurements are performed by applying an AC tip bias V$_{\rm tip}$ at a frequency of 50~kHz and an amplitude of V$_{\rm \emph{p}-p}$~=~1~V.
Lock in detection of the analog output of the IQ mixer yields the differential signals S$'_{\rm G}$~$\equiv$~dG/dV$_{\rm tip}$ and S$'_{\rm C}$~$\equiv$~dC/dV$_{\rm tip}$ \cite{berweger15,huber12}.

We note that in principle the differential relationship between $\tilde{\rm S}'$ and $\tilde{\rm S}$ suggests that a single measurement would be sufficient and the other data channel can be determined via numerical integration or differentiation at each spatial pixel.
However, differential operations significantly increase the noise and are highly susceptible to sample drift.
In practice, due to the higher contrast in $\tilde{\rm S}'$ integration is preferable, but large backgate voltages would be required to establish the constant of integration.

In order to minimize the influence of device hysteresis on our measurements we sequentially perform SMM imaging at each decreasing value of V$_{\rm BG}$ followed by a current-voltage (IV) sweep. 
We find a high degree of reproducibility when repeating V$_{\rm BG}$-dependent image sequences.
The device global conductivity minimum extracted from the sequence of IV curves corresponds to the condition where $\sigma_e$~=~$\sigma_h$, but since $\mu_e$~$\approx$~$\mu_h$ \cite{wang18,qiu18} this also corresponds to the approximate carrier equivalence (charge neutrality) point.

*Mention of commercial products is for informational purposes only, it does not imply NIST's recommendation or endorsement.

\section{Acknowledgements}
We would like to thank Kevin J. Coakley for valuable discussions.
The work at Purdue University is partly supported by NSF grant no. CMMI-1663214 and the Army Research Office under grant nos. W911NF-15-1-0574 and W911NF-17-1-0573

\bibliography{Te}

\end{document}

Here we provide the full data sets for the devices shown in the manuscript, as well as an additional thin device. 

\begin{figure}[!htb]%%%%%%%%%%%%%%%%%%%%%%%%
\includegraphics[width=.9 \textwidth]{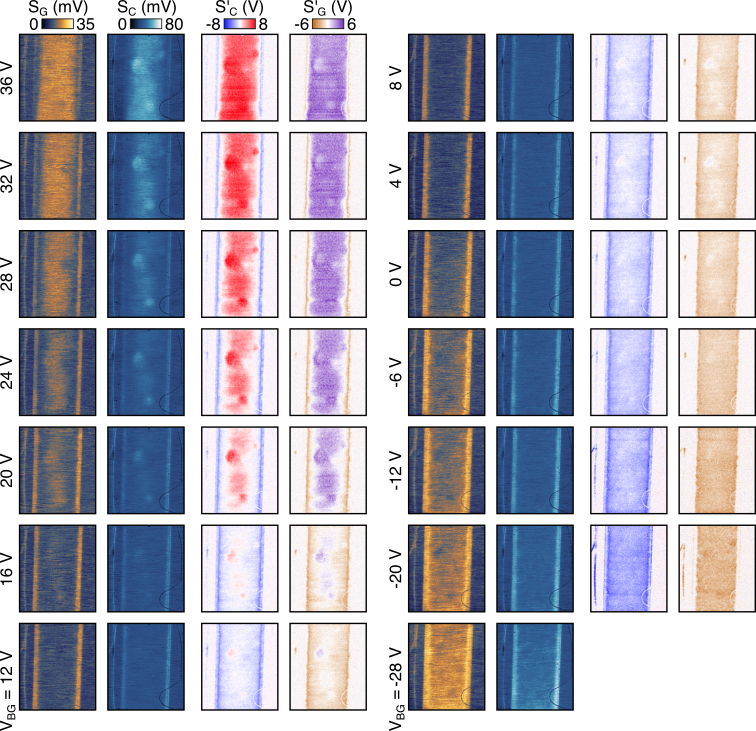}
\caption{
Full data set to device shown in Figs. 2 and 3 of the manuscript.
}
\label{supp1}
\end{figure}

\begin{figure}[!htb]%%%%%%%%%%%%%%%%%%%%%%%%
\includegraphics[width=.9\textwidth]{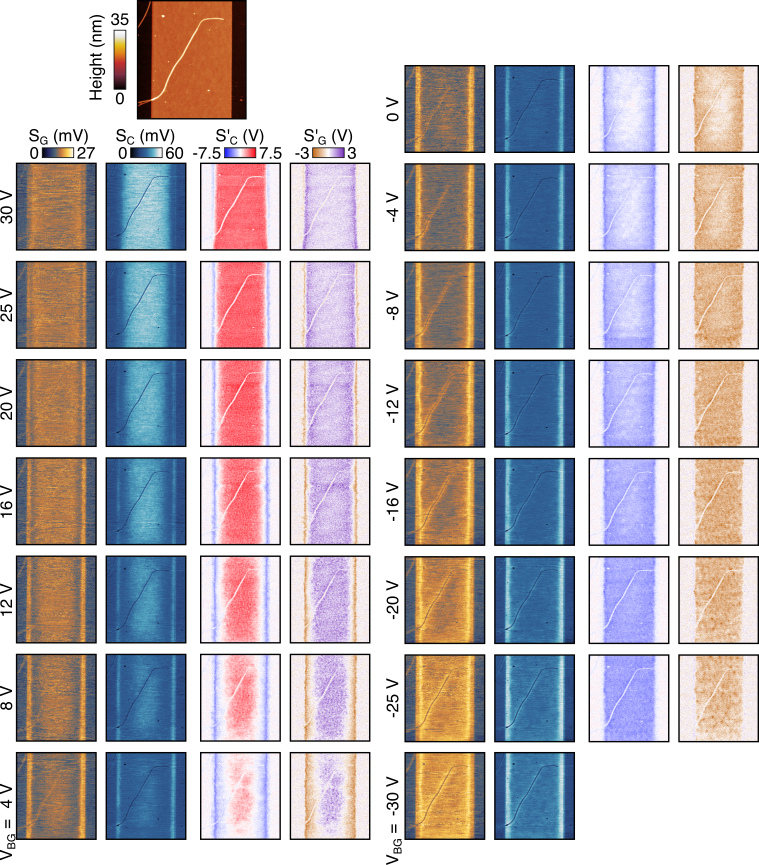}
\caption{
Full data set for an additional device.
This device is similar to the one shown in Figs. 2 and 3 of the manuscript with similar thickness and overall size. 
For this particular device the device carrier equivalence point is measured to be at V$_{\rm BG}$~=~0~V.
}
\label{supp2}
\end{figure}

Shown in Fig.~\ref{supp1} is the full data set for the device shown in Figs. 2 and 3 of the manuscript.
This device is fabricated on 90~nm of SiO$_2$ with a 10~nm conformal overcoat of alumina (Al$_2$O$_3$) deposited to prevent electrical discharge from the tip to the device.
For this particular device the global carrier equivalence point is observed to occur at V$_{\rm BG}$~=~20~V.
Shown in Fig.~\ref{supp2} is the full data set for a device fabricated under identical conditions as the one in Fig.~\ref{supp1} with a thickness of $\sim$8~nm. 
The device carrier equivalence point is measured to be at V$_{\rm BG}$~=~0~V. 
With both of these devices Te filaments are seen next to as well as on top of the devices, though the electrical impact of these appears to be weak.

%%%%%%%%%%%%%%%%%%%%%%%%

%%%%%%%%%%%%%%%%%%%%%%%%

\begin{figure}[!htb]%%%%%%%%%%%%%%%%%%%%%%%%
\includegraphics[width=.7\textwidth]{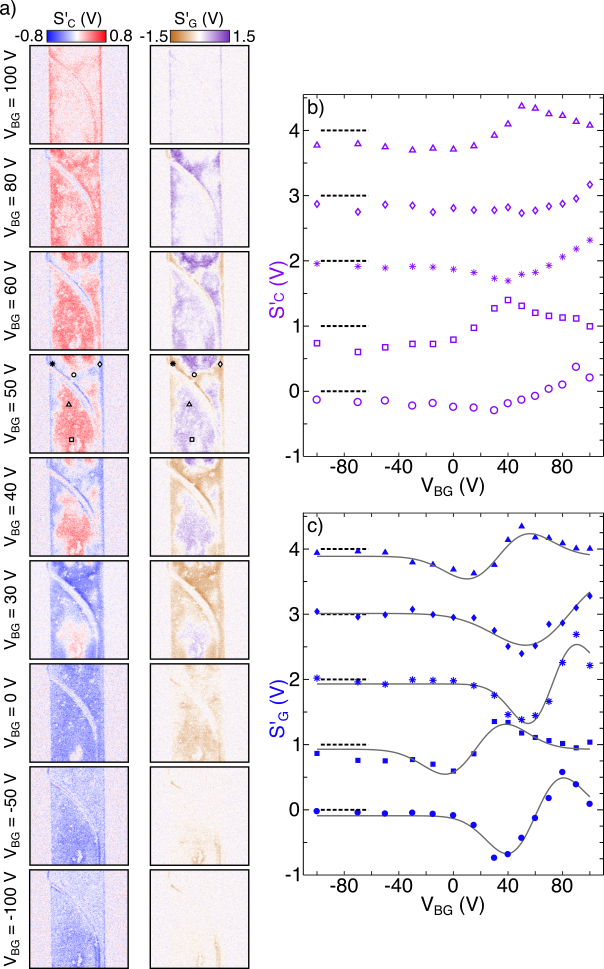}
\caption{
Full data set for the device shown in Fig.~4 of the manuscript.
}
\label{supp3}
\end{figure}

Shown in Fig.~\ref{supp3}a is the complete data set for the device shown in Fig.~4 of the manuscript. 
This device was fabricated on 300~nm oxide with a 10~nm conformal overcoat of hafnia (HfO$_2$).
Due to the thicker oxide, larger backgate voltages are required for effective gating, the device can readily sustain values of up to V$_{\rm BG}$~=~$\pm$ 100~V, and the global carrier equivalence point is seen to occur at V$_{\rm BG}$~=~60~V.
Importantly, based on the AFM topography we infer that the filament is embedded beneath the device, significantly affecting the spatial carrier distribution.
For this device only S$'_{\rm C}$ and S$'_{\rm G}$ were measured.

%%%%%%%%%%%%%%%%%%%%%%%%

Shown in Fig.~\ref{supp3}b and c are the V$_{\rm BG}$-dependent values of S$'_{\rm C}$ and S$'_{\rm G}$.
We see that the zero-crossing of interest, representing the local carrier equivalence point is the same for both data channels, as expected. 
In order to optimally extract the zero-crossing and enable extrapolation to values outside of the measured range, we observe that an excellent phenomenological fit to S$'_{\rm G}$ can be obtained of the functional form:
\begin{equation}
\frac{dG}{dV}(V_{\rm BG})=A_o +A(V_o-V_{\rm BG}) e^{-\frac{(V_o-V_{\rm BG})^2}{2\sigma^2}}.
\label{eq1}
\end{equation}
with the fit to each curve shown by the gray lines in Fig.~\ref{supp3}c). 
The carrier equivalence point voltages shown in Fig.~4d) of the manuscript are given by the values of V$_o$ extracted at each spatial pixel by fitting Eq.~\ref{eq1}.

\begin{figure}[h]%%%%%%%%%%%%%%%%%%%%%%%%
\includegraphics[width=.7\textwidth]{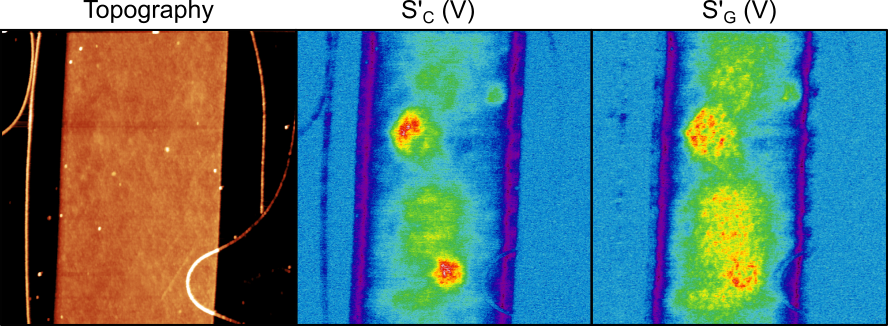}
\caption{
Topography and corresponding microwave signal for a Te device, where the color bars are scaled to maximize the visibility of small variations in all signal channels within the crystal.
}
\label{supp4}
\end{figure}%%%%%%%%%%%%%%%%%%%%%%%%

Lastly, we show a high resolution image of a device. 
Shown in Fig.~\ref{supp4} are high resolution images of the device shown in Fig.~\ref{supp1} and Figs. 2 and 3 with V$_{\rm BG}$~=~16~V. 
The colorbars for all images have been scaled to optimize the visibility of small features within the tellurene crystal. 
In particular, in the topography we see striations that align along the length of the crystal, which also correlate with visible variations in the microwave signal. 
As S$'_{\rm C}$ and S$'_{\rm G}$ are free of topographic artifacts, this signal arises from variations in electronic properties that correlate with physical structure.

%\bibliography{Te} 